# Predicting skull lesions after clinical transcranial MRI-guided focused ultrasound with acoustic and thermal simulations

Nathan McDannold, P. Jason White, and Rees Cosgrove

*Abstract*— Transcranial MRI-guided focused ultrasound (TcMRgFUS) thermal ablation is a noninvasive functional neurosurgery technique. Previous reports have shown that bone marrow damage in the skull can occur at high acoustic energies. While this damage is asymptomatic, it would be desirable to avoid it. Here we examined whether acoustic and thermal simulations can predict where the thermal lesions in the skull might occur. Post-treatment imaging was obtained at 3-15 months after 40 clinical TcMRgFUS procedures, and skull lesions were observed after 16/40 treatments. The presence for lesions was predicted by the acoustic energy with a threshold of 18.1-21.1 kJ (maximum acoustic energy) and 97-112 kJ (total acoustic energy). The size and degree of the lesions was not always predicted by the acoustic energy used during treatment alone. In contrast, the heating estimated by the acoustic and thermal simulations was predictive of the lesion extent. The lesions appeared in areas that were predicted to have high temperatures, and when thin slice T2-weighted imaging was obtained at 3-15 months, an excellent agreement was observed between the predictions and lesion locations. While more work is needed to validate the absolute temperatures measurements in and around the skull, being able to predict the locations and onset for skull lesions could allow for better distribution of the acoustic energy over the skull. Understanding skull absorption characteristics of TcMRgFUS could also be useful in optimizing transcranial focusing.

## I. Introduction

TRANSCRANIAL MRI-guided focused ultrasound (TcMRgFUS) is a noninvasive neurosurgical method that is approved for the treatment of essential tremor and tremor-dominant Parkinson's disease [1, 2], and it is being investigated for other functional neurosurgery applications [4-6]. The method uses a hemispherical phased array transducer to focus ultrasound energy through the human skull to thermally ablate small volumes in central brain locations. The magnitude and phase offsets for the transducer elements are manipulated to correct for aberrations of the ultrasound field induced by the skull bone and to normalize the acoustic intensity on the skull. These offsets are determined based on the geometry and density of the skull bone that are obtained from CT scans [7, 8]. The procedure is performed within an MRI scanner and guided by MR temperature imaging (MRTI) [10] and the patient's clinical response.

To avoid overheating the skull, scalp, and brain surface, a low acoustic frequency is used to minimize absorption in the bone, a hemispherical transducer design is employed to distribute the acoustic energy over a large surface area, and the head is actively cooled with chilled and circulating water. The acoustic energy needed to achieve an effective thermal dose at the focus varies substantially [12]. Depending on the geometry and acoustic properties of the skull, the absorbed energy can be sufficient to cause changes in the bone marrow. Schwartz et al. reported skull marrow changes following TcMRgFUS three months after sonication at acoustic energies of 15.5-25.8 kJ [13].

Damage to the marrow is asymptomatic and does not appear to pose a risk of harm for TcMRgFUS patients. Nevertheless, it could be desirable to avoid it and to predict where excessive heating might occur. Furthermore, observing where it occurs could improve our understanding of the acoustic propagation across the bone. This analysis could ultimately be used to optimize the aberration correction and perhaps devise new strategies to improve the size, shape, and obliquity of the focal lesion at the brain target.

The purpose of this study was to investigate the use of acoustic and thermal simulations to predict the onset and extent of skull damage after TcMRgFUS thermal ablation. Skull lesions were identified and segmented on T2-weighted MRI obtained 3-15 months after 40 clinical treatments and registered to a common reference frame. The locations of the skull lesions were then compared to the predicted heating based on acoustic simulations obtained in an earlier study [12]. Previous reports have studied skull heating during TcMRgFUS [14, 15]. Here we demonstrate that these simulations can accurately predict the skull lesion locations, providing initial clinical validation of these models.

This work was supported by the NIH under Grant R01EB025205.

N. McDannold is with the Radiology Department at The Brigham and Women's Hospital and Harvard Medical School, Boston, MA 02115 USA (e-mail: njm@ bwh.harvard.edu).

R. Cosgrove is with the Neurosurgery Department at The Brigham and Women's Hospital and Harvard Medical School, Boston, MA 02115 USA (e-mail: rcosgrove2@bwh.harvard.edu).

P.J. White is with the Radiology Department at The Brigham and Women's Hospital and Harvard Medical School, Boston, MA 02115 USA (e-mail: white@ bwh.harvard.edu).

## II. METHODS

### A. TcMRgFUS

The patients were treated using the ExAblate Neuro device (InSightec, Haifa, Israel), which uses a 1024-element phased array transducer to focus ultrasound through the human skull into central locations of the brain for thermal ablation. The elements have a central frequency of 660 kHz and are arranged in a 30 cm diameter hemisphere. The TcMRgFUS device is integrated into a 3T MRI scanner (GE750W, GE Healthcare, Milwaukee, WI). Imaging during the treatments was performed using the body coil. The transducer location within the MRI coordinate system was found using MRI tracking coils. The patient was placed in a stereotactic frame that was attached to the MRI table. The space between the scalp and the transducer face was filled with degassed water that was chilled to 15°C and circulated between sonications. A flexible membrane placed around the patient's shaved head and attached to the face of the transducer contained the water.

Over a three-year period, we treated 61 patients at our institution. Overall, 59 patients were treated for essential tremor (thalamotomy) and two for Parkinson's disease (pallidotomy). MRI was obtained 3-15 months after TcMRgFUS in 40 patients.

### B. Imaging

Before treatment, a CT scan was obtained with a bone reconstruction kernel. On treatment day, neurosurgical planning was performed using a 3D FIESTA sequence. One day after treatment, an MRI exam was acquired on a 3T scanner at our institution. At 3-15 months after treatment, an additional MRI exam was obtained in the majority of patients at either 3T or 1.5T. All MRI exams included an axial T2-weighted sequence, which we used to detect treatment-induced changes in the skull. Details about the patients, treatment parameters, and post-FUS imaging are shown in Table S1.

### C. Registration

The transformation matrices describing the registration between the treatment planning images obtained with the patient in the TcMRgFUS device, the pre-treatment CT scan, and the reference frame of the transducer were obtained from treatment log files exported from the device. The registrations between the MRI obtained during treatment and that obtained 24 h and 3-15 months after FUS were created manually using 3D slicer (www.slicer.org). After registration, comparisons of axial T2-weighted MRI obtained at 24 h and 3-15 months later was used to identify skull damage. The extent of the lesions at 3-15 months was manually segmented by one author (NM). All imaging and the segmentations were then interpolated using MATLAB into the same 256×256×145 element matrix, which was in the reference frame of the phased array transducer. This matrix had dimensions of 30×30×17 cm; bicubic interpolation was used.

### D. Acoustic and thermal simulations

Acoustic simulations were performed in an earlier study [12] using k-Wave, an open source MATLAB toolbox [16]. We simulated the pressure field in a small volume for each transducer element separately (Fig. 1a). These simulations were performed in the coordinate system of the individual elements (elemental simulation matrix size: 44×44×492), and the resulting complex pressure fields were rotated and interpolated into the reference frame of the TcMRgFUS transducer and summed (Fig. 1b). In this study we interpolated to the same 256×256×145 element matrix that we used for the imaging data. We assumed the elements were 1 cm diameter pistons centered at the element locations provided by the manufacturer. The simulations were performed using a spatial resolution of 0.325 mm (wavelength/7). We used a perfectly matched layer with a size of 10 grid points and an attenuation of 2 Np per grid point. The acoustic simulations were run in parallel using a computing cluster. Parameters used for the acoustic and thermal simulations are shown in Table 1. We used the relationships between the skull density and the attenuation/sound speed found in our earlier study [12].

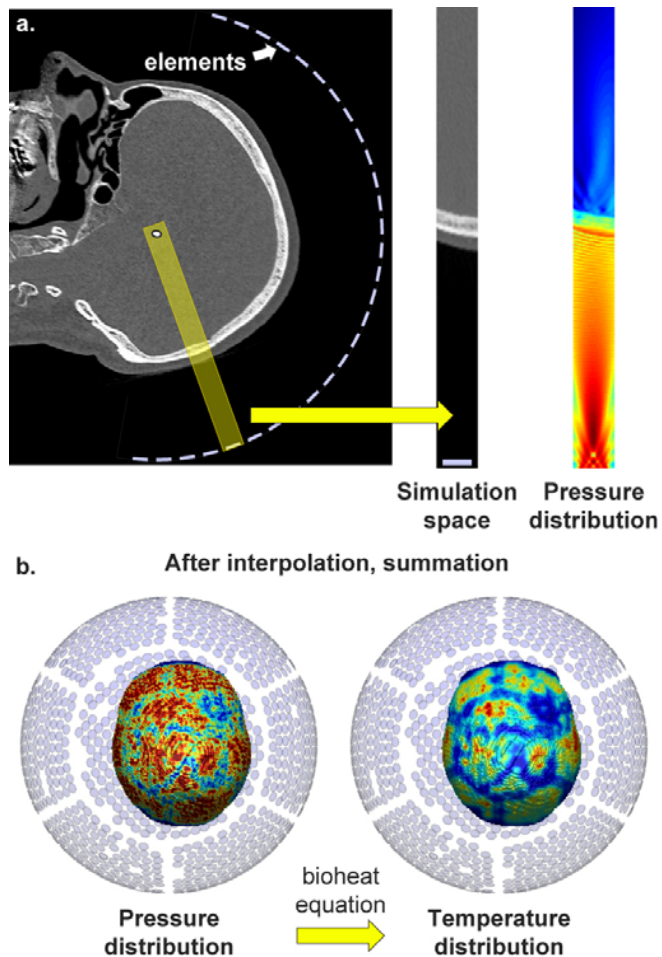

Fig. 1. Methods. (a) CT scan data (sagittal reconstruction) of a patient with the transducer elements superimposed. For each element, a small volume was selected. The CT data was used to estimate the acoustic parameters (density, sound speed, attenuation) that were inputs to the acoustic simulation. (b) After all elements were simulated, the resulting pressure fields were interpolated into a common reference framed and summed. This combined field was used with the bioheat equation to estimate the temperature rise.

TABLE I
ACOUSTIC AND THERMAL PARAMETERS USED IN THE NUMERICAL MODEL

| Parameter | Water | Brain | Skull |
| --- | --- | --- | --- |
| Density (mg/kg³) | 1000 | 1030[a] | from CT[d] |
| Sound speed (m/s) | 1500 | 1560[a] | from CT[d] |
| Attenuation at 660 kHz (Np/m) | 0 | 4.36[b] | from CT[d] |
| Absorption at 660 kHz (Np/m) | 0 | | 20.5 (cortical)[c] 41.0 (diploe)[c] |
| Thermal conductivity [W/(m·°K)] | - | 0.51[a] | 0.43[a] |
| Specific heat [J/(kg·°K)] | - | 3640[a] | 1440[a] |
| Perfusion coefficient (l/s) | - | 8.33E-03[a] | 3.33E-04 |

[a]ref. [3]; [b]ref. [9]; [c]ref. [11]; [d]see Ref [12] for details.

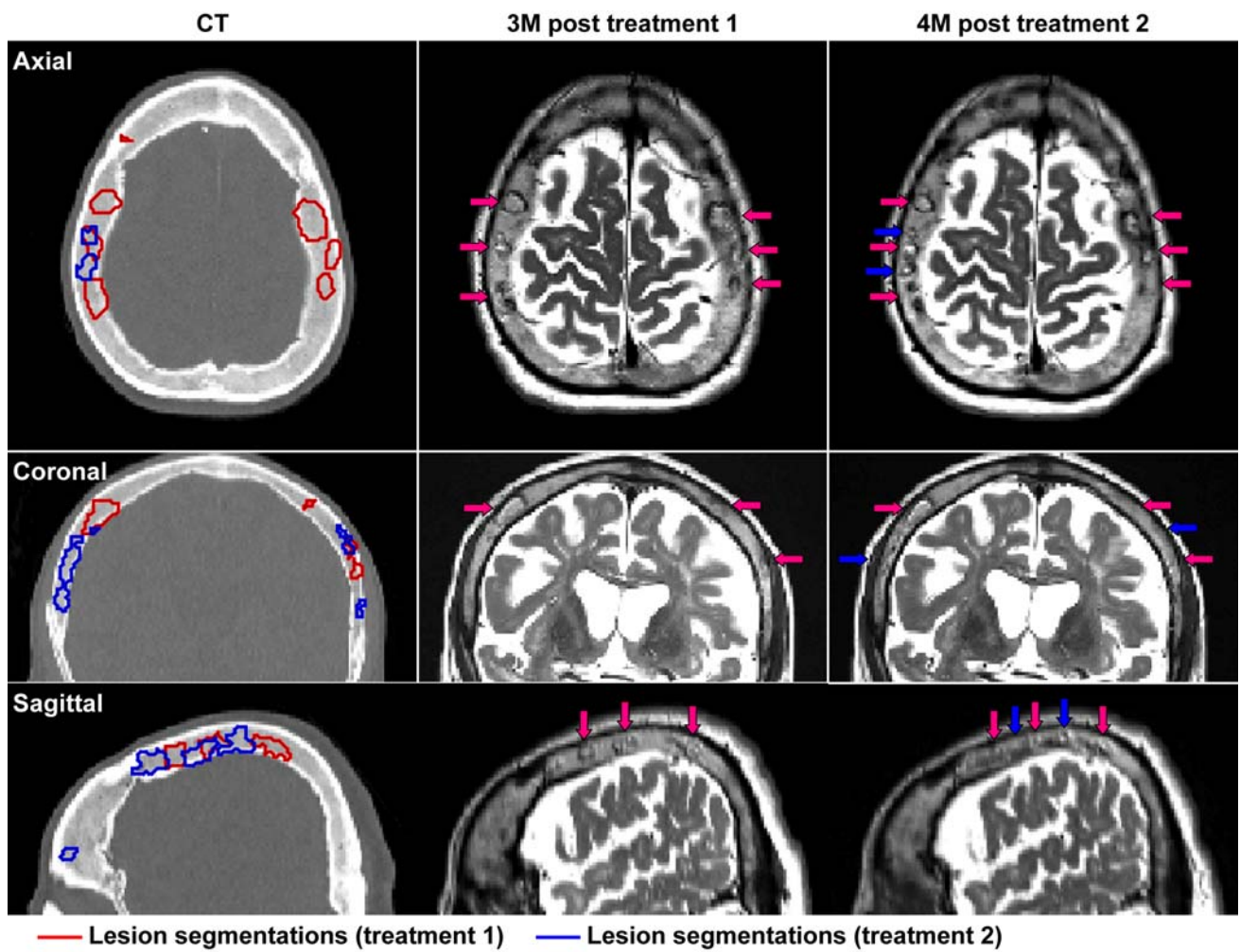

Fig. 2. Example skull lesions in an ET patient who was treated twice. Left: CT scan with the segmentations of the lesions superimposed. Center/right: T2-weighted MRI obtained 3-4 months after the two treatments. All images were registered and interpolated to a common coordinate system.

For each treatment, we used the phase and magnitude values for the transducer elements that were employed during the sonication with the highest acoustic energy. We estimated the heating in and around the skull with the bioheat equation [17]. For the thermal simulations, we used absorption coefficients of 20.5 and 41.0 Np/m for cortical and trabecular bone, respectively. These values were based on the study by Pinton et al., who used simulations and high-resolution micro-CT to estimate the longitudinal and shear absorption coefficients of the human skull [11]. We used the shear absorption coefficient for the trabecular bone based on an assumption that the complex bony structure in the diploe would be higher because of shear absorption. We previously identified the location of the inner and outer table for each voxel in the simulations [12]. Voxels between these two points with a density of 2000 kg/m³ or less was characterized as trabecular bone.

Because of uncertainty in absorption coefficient, the absolute temperature predicted by the thermal simulations were not expected to be precisely accurate, but instead were intended to provide a relative prediction of the pattern and severity of skull heating. The head was assumed to be at a uniform temperature of 37°C before sonication. The surrounding water was assumed to be 15°C. We did not simulate the cooling of the head before sonication due to the chilled and circulating water.

### E. Data analysis

To compare the location and extent of the skull lesions to the simulated heating, we created maps of maximum temperature rise that were superimposed on the skull surface obtained from the CT scan. We further superimposed isosurfaces of the skull lesions segmentations on these maps. Agreement was assessed qualitatively. The presence (or lack) of skull lesions was plotted as a function of the maximum acoustic energy used during each treatment as well as the total applied energy. These data were fit using probit regression to estimate the probability for skull lesions along with 95% confidence intervals. All data analyses were performed in MATLAB.

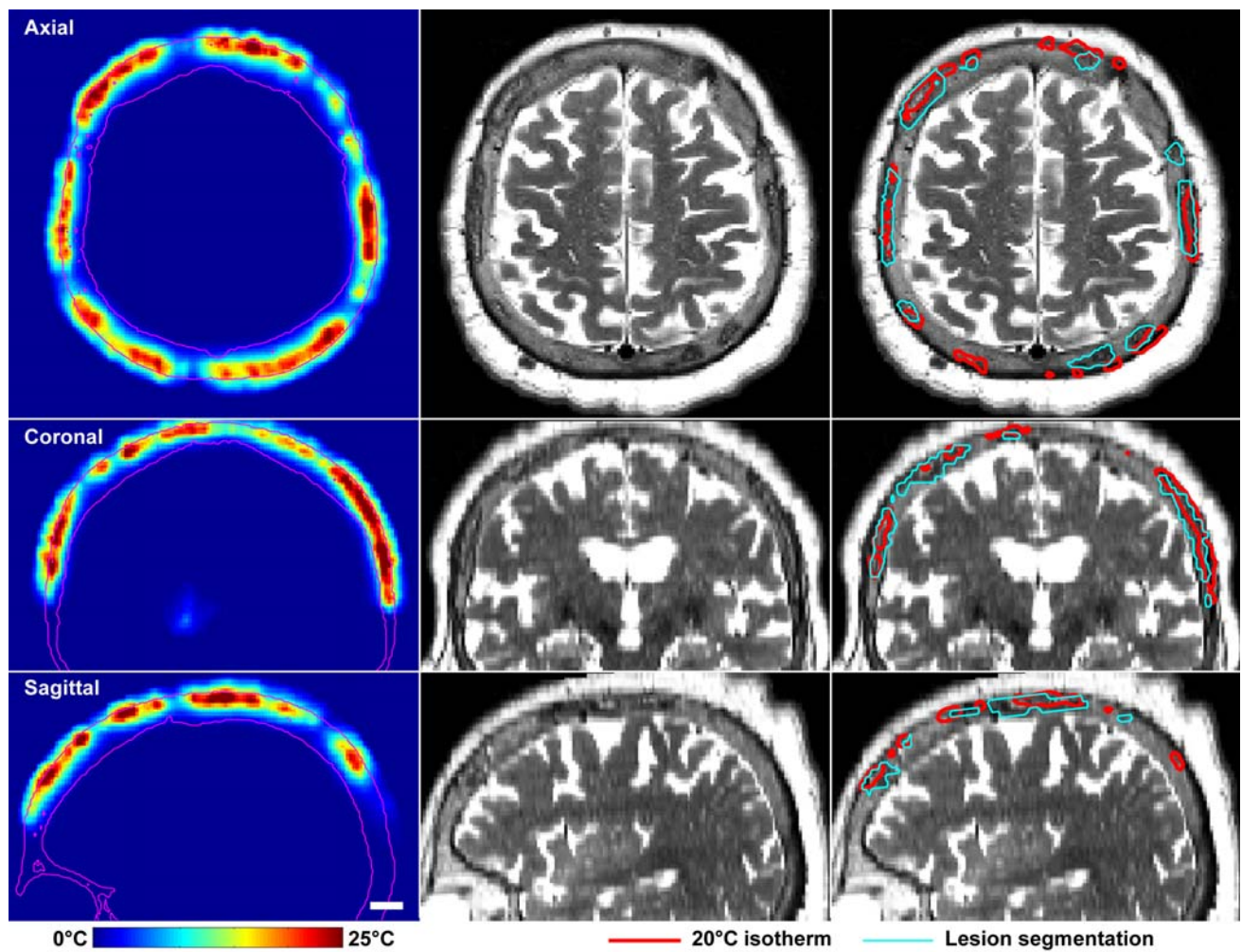

Fig. 3. Simulated skull heating and lesions. Left: Simulated heating in treatment 12. The outline of the skull from the CT scan is superimposed. Middle: T2-weighted MRI obtained four months after TcMRgFUS. Right: The same images with segmentations of the skull lesions and 55°C isotherms superimposed. All images and segmentations were registered and interpolated to the same reference frame.

### III. RESULTS

T2-weighted MRI was obtained 3-15 months after 40 TcMRgFUS treatments in 38 patients; two patients received two treatments. Obvious skull lesions were detected at 3-15 months after TcMRgFUS in 40% (16/40) of the treatments. The damage was not evident at 24 h after treatment in any patient. The lesions appeared as an isointense or heterogeneous core surrounded by a hypointense rim, or for smaller lesions, simply a hypointense area.

Fig. 2 shows the examples of skull damage in a patient that received two TcMRgFUS treatments. At three months after the first treatment, numerous damaged regions were observed. At four months after the second treatment, the lesions produced by the first treatment were still evident. New lesions were detected at this time on the temples and in an anterior location in addition to those produced by the first treatment.

Fig. 3 shows another example of the skull lesions in a pa-tient observed four months after TcMRgFUS, here along with heating predicted by the simulations. In the simulation, numerous spots were predicted with high temperatures. Comparison of isotherms with the lesion segmentations reveal that the location of the lesions was consistent with the predictions of the simulation. The simulation also predicted low-level heating just inside the skull bone in the subdural space; such heating was observed in MRTI (Fig. 4).

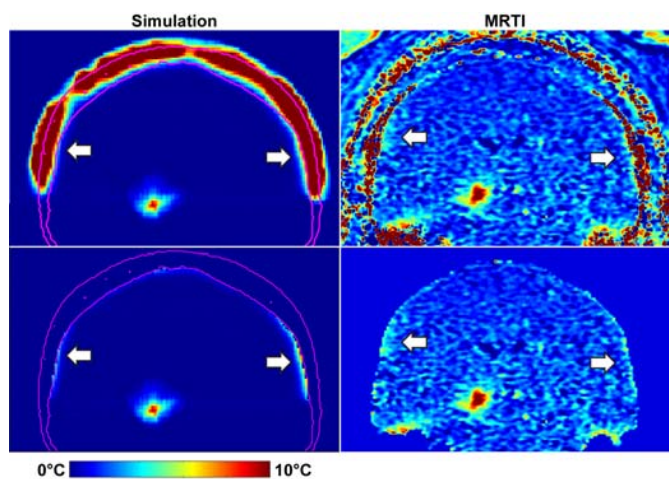

Fig. 4. Coronal MRTI obtained during a sonication in treatment 12. Heating just inside the skull bone (arrows) was observed in MRTI and predicted by the simulation. Note that the bone in the MRTI does not appear to be heated due to the lack of MRI signal. The images on the bottom have masked out everything except for the subdural space to highlight the low-level heating.

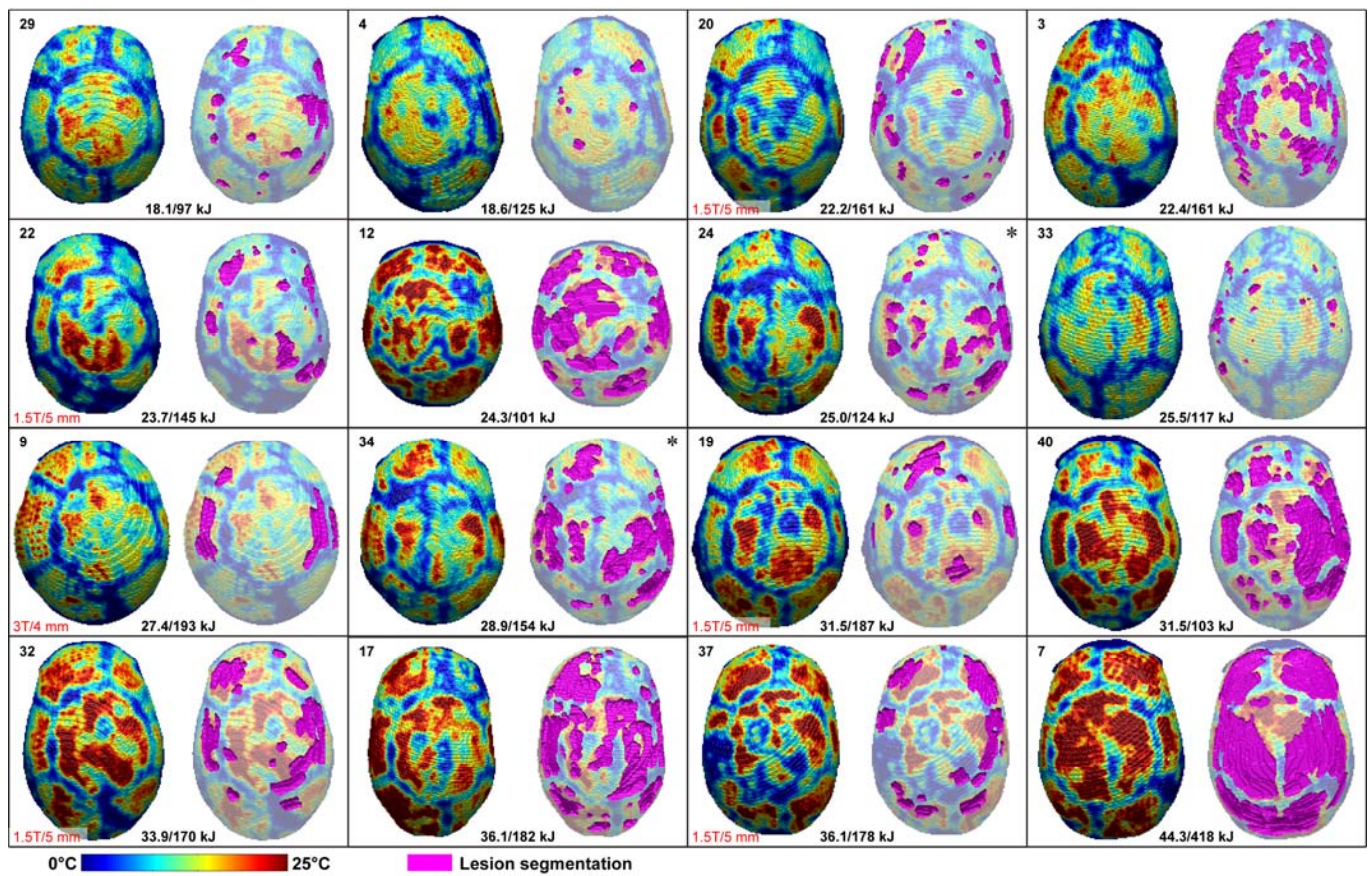

Fig. 5. Skull heating and lesions. Maps of the outer skull surface are shown for 16 patients where skull lesions were observed 3-15 months after TcMRgFUS. The maps are color-coded with the maximum temperature projections estimated by the simulations. The maps on the right show the locations of the skull lesions in magenta. In most cases, the locations of the skull lesions corresponded to regions where high temperatures were predicted by simulation. Except where noted, the post-FUS MRI was obtained at 3T with a 2 mm slice thickness. When thicker slices were used, high temperatures were often predicted in dorsal skull regions where lesions were not observed, suggesting perhaps lesions were present but not observed in MRI. The maps are sorted by the maximum acoustic energy used during the treatments. The treatment number and the maximum/total acoustic energy are shown; one patient received two treatments (*) 3 months apart. Treatment 9 was a pallidotomy for Parkinson's disease; others were thalamotomy for essential tremor.

There was excellent agreement between the simulations and the skull lesion locations in all of the patients. Fig. 5 shows a 3D rendering of the outer skull surface color-coded by the maximum temperature rise in the bone and the lesion locations superimposed. In almost every case, the lesions occurred in regions where the temperature was predicted to be elevated. However, there were also numerous locations where the simulations predicted high temperatures where lesions were not observed. In several cases, these heated regions were in dorsal portions of the skull and occurred when the 3-15 month MRI was obtained with thick slices, suggesting perhaps that lesions were present but not detected.

Fig. 6 shows the simulated heating for the 24 patients where lesions were not detected. Overall, the predicted heating was substantially less in most of those patients. However, four treatments had regions where the simulated heating was comparable to those where lesions were observed in other patients. Three of these cases had thick slices in the post-treatment MRI which may have failed to detect more subtle or smaller lesions.

Both the maximum and total energy used during each TcMRgFUS treatment could reliably predict the onset for skull damage (Fig. 7). The threshold for damage was between 18.1-21.1 kJ (maximum energy applied) and 97-122 kJ (total energy applied). In contrast, the extent of the skull lesions at 3-15 months was not always well-predicted by the energy alone. The maps in Fig. 5 are sorted by the peak sonication energy used during each treatment. While overall the extent of the lesions increased as the energy increased (Fig. 8a), there were several cases that did not fit this trend. For example, treatment number 12 had extensive lesions after TcMRgFUS where the peak and total acoustic energy was 24.3 and 101 kJ, respectively. Treatment number 33, in contrast, had only small lesions after sonications at higher energies. The simulations predicted substantially higher skull temperatures in treatment 12 and lower temperatures in treatment 33.

The volume of the heated regions estimated by the simulations was correlated with the lesion volume (Fig. 8b). The mean estimated temperature in the skull in the areas identified as skull lesions was 54.2 ± 7.0 °C. Areas not identified as lesions had a significantly (P<0.001) lower mean temperature of 44.6 ± 6.6 °C.

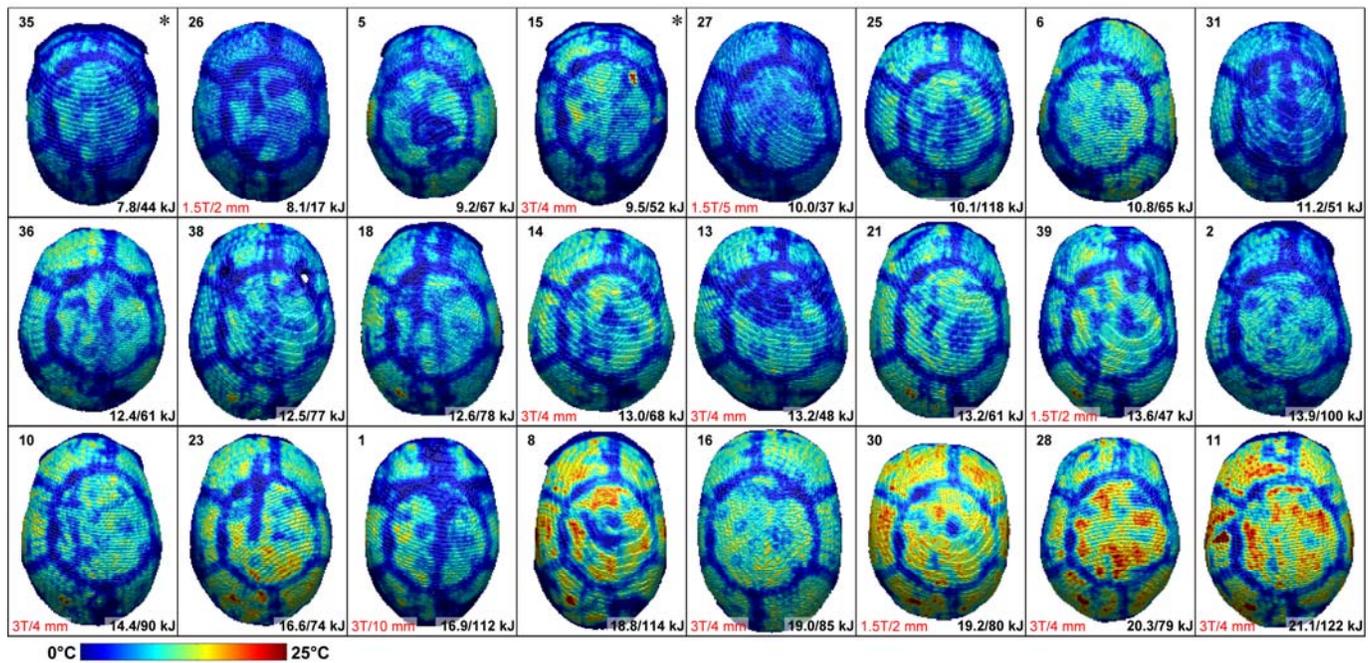

Fig. 6. Skull heating in 24 treatments where skull lesions were not observed. The maps are color coded with the maximum temperature predicted by simulations of the sonication with the maximum energy for each treatment. Except where noted, the post-FUS MRI was obtained at 3T with a 2 mm slice thickness. The maps are sorted by the maximum acoustic energy used during each treatment; the treatment number along with the maximum and total acoustic energy used are noted. One patient received two treatments (*) 10 months apart. Treatment 10 was a pallidotomy for Parkinson's disease; others were thalamotomies for essential tremor.

## IV. DISCUSSION

The acoustic energy used during TcMRgFUS was predictive of whether or not skull lesions were observed at 3-15 months after treatment, and the threshold for skull damage (18.1-21.1 kJ) is consistent with earlier work (15.5-25.8 kJ) [13]. However, the acoustic energy was not always predictive of the extent of the damage. In contrast, the acoustic and thermal simulations together did a good job in predicting both the location and extent of the resulting lesions. If the simulations could be performed before treatment, the absorption and subsequent heating could be taken into account, potentially allowing one to change how the energy is distributed over the head to better normalize the heating. With enough confidence in the simulations, one could potentially redistribute the energy delivered into the bone to better optimize the shape and size of the focus while maintaining a safe exposure level overall.

To use the simulations in this way, we need to validate the absolute temperature predictions. An important factor that was not investigated here is the effects of the active cooling on the baseline temperature of the scalp, bone, and brain surface. Methods developed to model hypothermia for the treatment of stroke [18] might be useful for such simulations.

This study had several other limitations. The MRI protocol at 3-15 months varied, and it appeared that that some lesions may have been missed when thicker imaging slices were used. It is also possible that some skull damage was resolved by the time that the imaging was obtained. Furthermore, since cortical bone does not have MRI signal in standard pulse sequences, any potential damage there was undetectable. It would be interesting to try ultra-short TE MRI [19] to see if any changes are evident in cortical bone. Obtaining CT scans after TcMRgFUS or contrast-enhanced MRI could also be revealing.

The simulations also had several limitations. They were previously obtained with the heating at the focus in mind. Each element was individually simulated over a small volume and summed. While around the focus the total combined field could be fully reconstructed, the dimensions of the individual elemental simulations may have been insufficient to provide full coverage at the skull, leading to an underestimate of the total field. However the main contribution from each element was included. The accumulated effects of multiple sonications were not taken into account. We also did not include shear mode conversion in the acoustic simulations. However, the TcMRgFUS system disabled elements with external angles greater than 25°, so such effects may not be significant. We also likely did not model absorption correctly in the diploe in the thermal modeling. More work is needed to understand the role of these factors along with the active cooling so that the absolute temperature in the bone can be accurately predicted. Despite these limitations, the relative heating distributions appears to reliably predict the locations of skull lesions post TcMRgFUS treatment.

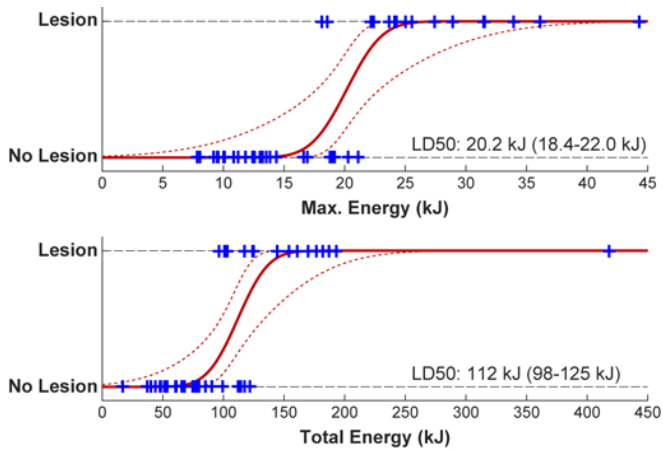

Fig. 7. Presence or lack of skull lesions plotted as a function of the maximum and total acoustic energy used during treatment. The probability for lesions is shown along with 95% CI.

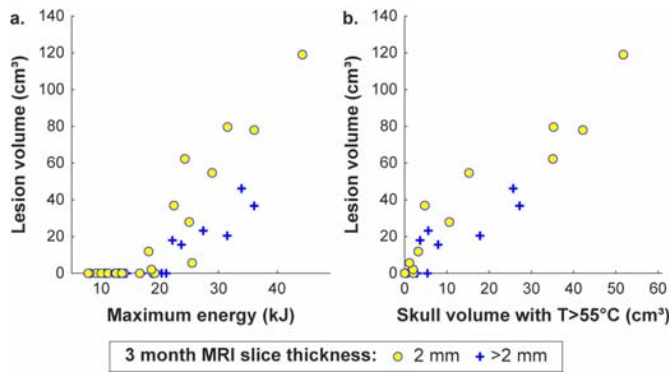

Fig. 7. Lesion volume plotted as a function of the maximum energy used during treatment and the volume of the skull that was estimated to have heated to 55°C or more.

## V. Acknowledgements

Portions of this research were conducted on the O2 High Performance Compute Cluster, supported by the Research Computing Group at Harvard Medical School. See http://rc.hms.harvard.edu for more information.